\documentclass{article}

\usepackage{arxiv}

\usepackage[utf8]{inputenc} 
\usepackage[T1]{fontenc}    
\usepackage{hyperref}       
\usepackage{url}            
\usepackage{booktabs}       
\usepackage{amsfonts}       
\usepackage{nicefrac}       
\usepackage{microtype}      
\usepackage{lipsum}
\usepackage{graphicx}
\usepackage{amsmath}
\graphicspath{ {./images/} }

\title{Network Sampling: An Overview and Comparative Analysis}

\author{
Quoc Chuong Nguyen \\
Department of Methematics\\
State University of New York at Buffalo\\
New York 14260, the Unites States \\
\texttt{quocchuo@buffalo.edu} \\
}

\begin{document}
\maketitle
\begin{abstract}
Network sampling is a crucial technique for analyzing large or partially observable networks. However, the effectiveness of different sampling methods can vary significantly depending on the context. In this study, we empirically compare representative methods from three main categories: node-based, edge-based, and exploration-based sampling. We used two real-world datasets for our analysis: a scientific collaboration network and a temporal message-sending network.  Our results indicate that no single sampling method consistently outperforms the others in both datasets. Although advanced methods tend to provide better accuracy on static networks, they often perform poorly on temporal networks, where simpler techniques can be more effective. These findings suggest that the best sampling strategy depends not only on the structural characteristics of the network but also on the specific metrics that need to be preserved or analyzed.  Our work offers practical insights for researchers in choosing sampling approaches that are tailored to different types of networks and analytical objectives.
\end{abstract}


\section{Introduction}
\begin{figure}[htp!]
\centering
\includegraphics[width=\linewidth]{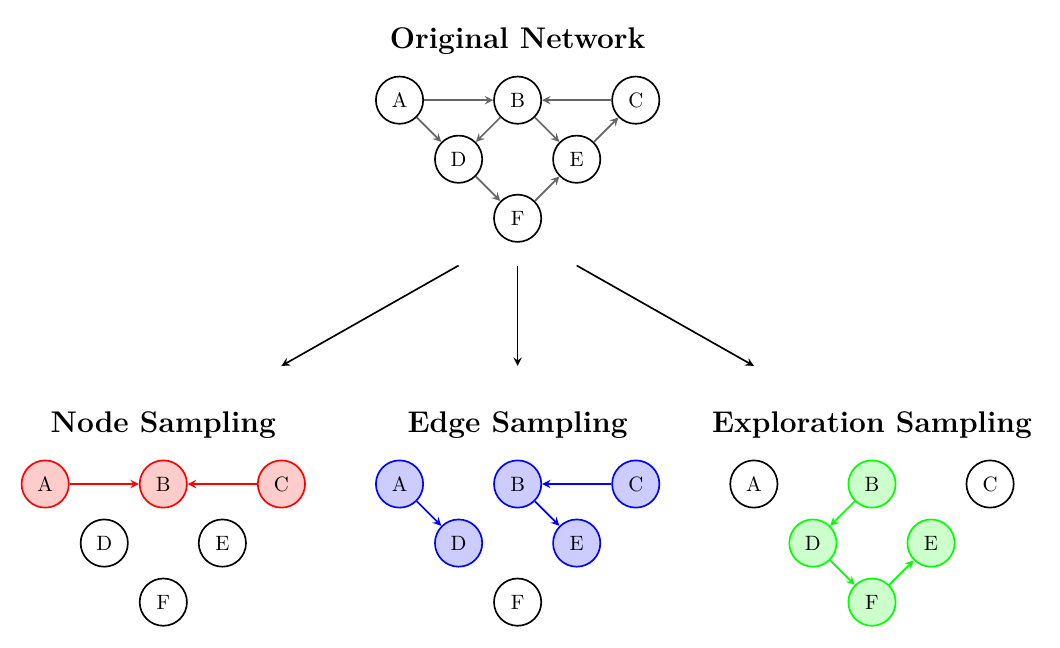}
\caption{Network sampling methodologies can be categorized into three primary approaches: node-based, edge-based, and exploration-based sampling. Node-based methods select nodes either uniformly or based on specific attributes. These methods often preserve properties at the node level. Edge-based techniques sample edges directly, which can better maintain connectivity patterns within the network. However, these methods may distort the distribution of nodes. Exploration-based methods, such as random walks or snowball sampling, traverse the network dynamically. These approaches are especially suited for large-scale or partially observable networks.}
\label{fig:method}
\end{figure}

The study of complex networks provides a fundamental understanding of multiple systems such as social interactions and biological processes, as well as information dissemination. Access limitations and the large dynamic scale of real-world networks prompt researchers to employ sampling methods to study network structures. The selection of sampling approaches plays a critical role because it greatly influences the accuracy of analysis techniques such as community detection and centrality estimation. Recent works \cite{doi:10.1177/0081175012461248,AIROLDI2011506,NBERw25270,https://doi.org/10.1002/mpr.2034,s21051905} emphasize that sampling strategies vary significantly in performance depending on network characteristics and analysis goals. As shown in Fig. \ref{fig:method}, sampling methods are generally categorized into three main classes:
\begin{itemize}
    \item \textbf{Node-based sampling} involves selecting nodes either uniformly at random or based on specific attributes, such as degree or activity level. This method is computationally efficient and commonly used in large-scale studies \cite{Ahmed2011NetworkSV,10.1145/3488560.3498383,MyakushinaExploringST,doi:10.1073/pnas.0501179102,PhysRevE.64.046135,10.1145/1150402.1150479}. However, it often fails to capture important global structural properties, including connectivity and clustering.
    \item \textbf{Edge-based sampling}, on the other hand, focuses on selecting edges either uniformly or according to their weights. These methods tend to better preserve structural patterns, including degree distribution and assortativity \cite{JMLR:v22:18-240,5961350,10192005,Jiao2024,10.1007/11422778_27,10.1145/2601438}. However, they may introduce bias into the sampled node set and distort centrality measures.
    \item \textbf{Exploration-based sampling} simulates network traversal through methods such as random walks, Metropolis-Hastings walks, or snowball sampling. This category is particularly effective for large, evolving, or partially observable networks \cite{5462078,4781123,Lawler1999,10.1214/aoms/1177705148,10.1145/1879141.1879192,10.1145/2505515.2505618,BLAGUS2017136}. However, it can be sensitive to the choice of initial seeds and can lead to an overrepresentation of densely connected regions.
\end{itemize}

Despite the increasing amount of work on the subject, there is still no agreement on which method performs best for various network types and analytical goals. In this study, we conduct a comparative evaluation of representative sampling methods from each category using two real-world datasets: a static scientific collaboration network and a temporal message-sending network. We examine each method's ability to maintain key structural properties, including degree distribution, clustering coefficient, component sizes, and temporal dynamics.

Our findings indicate that no sampling method consistently outperforms others across different datasets and metrics. Notably, while advanced techniques such as exploration-based sampling show strong performance in static networks, simpler strategies like uniform node or edge sampling tend to produce better results in temporal contexts. This variability highlights the importance of customizing sampling strategies based on both the network structure and the specific metrics being analyzed. Our work offers practical insights for researchers dealing with the trade-offs involved in network sampling and emphasizes the need for context-aware approaches in both static and dynamic network environments.

\section{Materials and Methods}

All the networks used in this study are undirected and unweighted. Depending on the characteristics we aim to analyze, we use two networks: the CA-HepTH dataset as the static network and the CollegeMsg dataset as the temporal and multiplex network.

\subsection{Definition of network sampling} 
Network sampling is the process of selecting a subset of nodes, edges, or paths from a large or partially observable network to facilitate analysis when accessing or processing the entire network is impractical. This approach is often necessary due to limitations in data availability, computational resources, or the dynamic nature of the network, particularly given the significant increase in the size of many practical networks today \cite{Touwen2024,doi:10.1073/pnas.2018994118,10204402, 10.3389/fdata.2022.797584,doi:10.1098/rspb.2011.1959}. Sampling allows for the efficient approximation of structural properties and metrics \cite{PhysRevE.73.016102,Ahmed_Neville_Kompella_2021,https://doi.org/10.1111/oik.08650,Smith2013,Zhang2025}, which are essential for tasks such as link prediction, community detection, and modeling spreading processes.

Formally, given a graph \( G = (V, E) \), where \( V \) is the set of nodes and \( E \subseteq V \times V \) is the set of edges, a sampling method produces a subgraph \( G_s = (V_s, E_s) \), where \( V_s \subseteq V \) and \( E_s \subseteq E \cap (V_s \times V_s) \). The goal is for \( G_s \) to preserve key properties of \( G \), such as degree distribution, clustering coefficient, or shortest-path length, within acceptable error bounds. Different sampling strategies; such as node-based, edge-based, and exploration-based approaches; each come with unique trade-offs in capturing different network properties. When selecting a specific method, it is crucial to consider both the structure of the network \( G \) and the specific metrics of interest. Sampling that does not align with these factors may lead to biased or misleading conclusions about the original network.

Our goal is to conduct a statistical comparison of sampling methods applied to real networks in order to better understand the advantages and disadvantages of each method. Despite the complexity of these methods, we will demonstrate that even in the case of small networks, more complex methods are not always superior. The choice of method significantly depends on the metrics we wish to extract from the original network.

\subsection{Static Network}
\label{sec:static}
\begin{figure}[tbhp]
\centering
\includegraphics[width=0.99\linewidth]{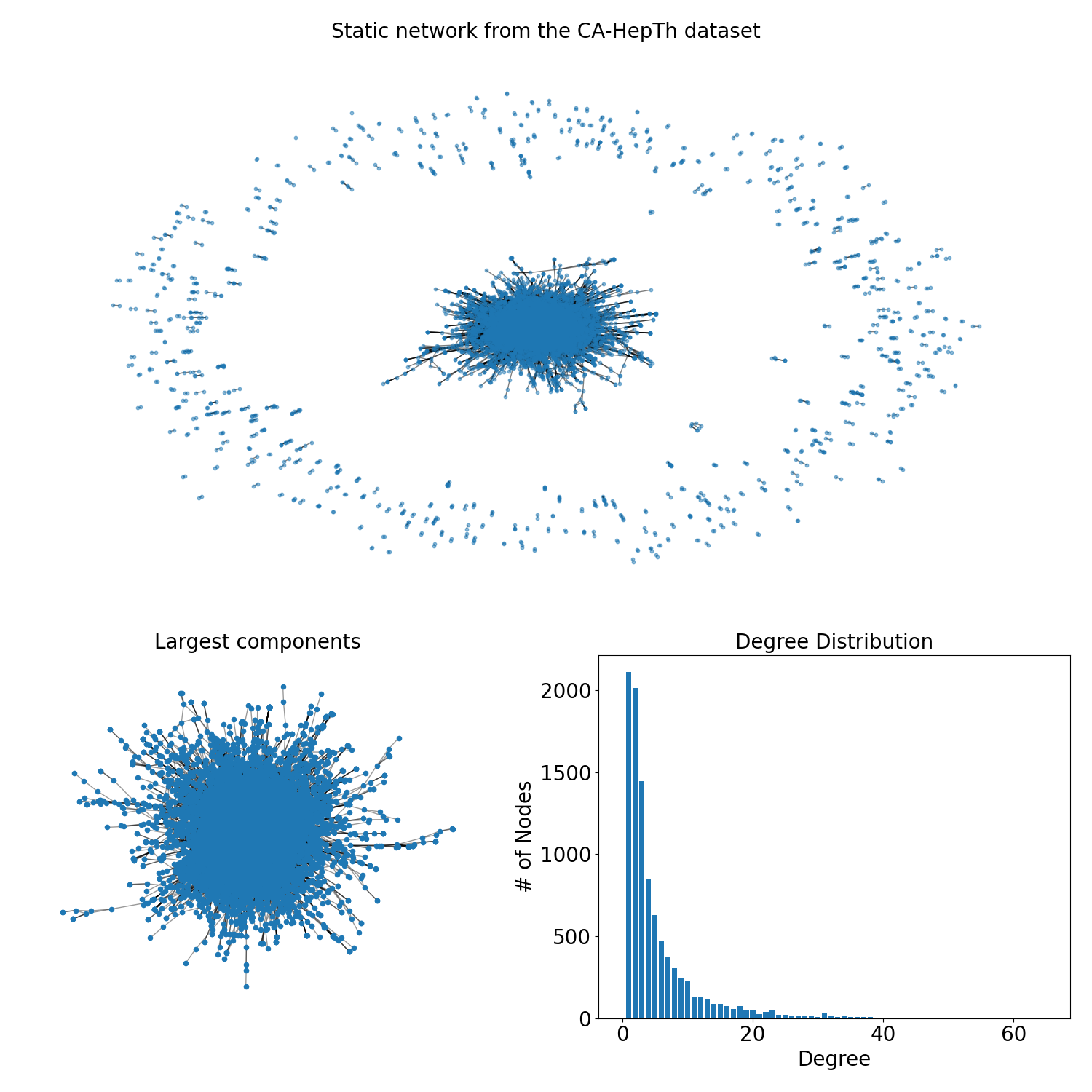}
\caption{The collaboration network, after the removal of loops (self-edges), contains 9,877 nodes and 25973 edges, with a clustering coefficient of 0.4714. There is one significantly large component, approximately 0.875 in size, which includes 8,638 nodes. The average shortest path length within this component is 5.95, indicating a small-world effect. Additionally, the degree distribution follows a power law, which is characteristic of a scale-free network.
}
\label{fig:dataset_static}
\end{figure}

The first dataset used in this project is the Arxiv HEP-TH \cite{10.1145/1217299.1217301} (High Energy Physics - Theory) collaboration network, obtained from the e-print service arXiv. It includes scientific collaborations among authors of papers submitted in the High Energy Physics - Theory category. In this network, if author \(i\) co-authored a paper with author \(j\), there is an undirected edge connecting \(i\) and \(j\). Consequently, this is an unweighted and undirected network. However, the original network contains several self-edges, which need to be removed to simplify the study, which is shown in Fig. \ref{fig:dataset_static}. A more detailed analysis of the collaboration network is shown in Table \ref{table:static}.

In the experimental setup, we compare six methods by evaluating the impact of subnetwork sizes on the approximated metrics. After generating 100 samples with one defined network size for each method, we will calculate the average metrics and compare them with the original network's metrics. The sampling methods in this experiment are listed as follows:

\begin{itemize}
    \item \textbf{Uniform Node Sampling (UNS)}: Randomly select nodes with equal probability \cite{doi:10.1073/pnas.0501179102}.
    \item \textbf{Weighted Node Sampling (WNS)}: Nodes selected based on their degree distribution \cite{PhysRevE.64.046135}.
    \item \textbf{Uniform Edge Sampling (UES)}: Randomly select edges with equal probability \cite{10.1007/11422778_27}.
    \item \textbf{Induced Edge Sampling (IES)}: Include both endpoints and all their mutual edges \cite{10.1145/2601438}.
    \item \textbf{Random Walk Sampling (RWS)}: Create an induced subgraph by randomly walking around \cite{5462078}.
    \item \textbf{Metropolis-Hastings Random Walk Sampling (MHRWS)}: Correct for bias in Random Walk Sampling by modifying transition probabilities \cite{4781123}.
    \item \textbf{Snowball Sampling (SS)}: Expand in waves from a seed node \cite{10.1214/aoms/1177705148}.
    \item \textbf{Breadth-First Search Sampling (BFS)}: Exploit the network by the Breadth-First Search algorithm.
\end{itemize}

\begin{table}[tbhp]
    \centering
    \begin{tabular}{|l l|} 
         \hline
         Metric & Value \\
         \hline
         Number of nodes & 9877 \\
         Number of edges & 25973 \\
         Average degree & 5.26 \\
         Clustering coefficient (C) & 0.4714 \\
         Largest component size (S) & 0.875 \\
         Average shortest path in S & 5.95 \\
         \hline
    \end{tabular}
    \caption{\normalfont A detailed analysis of the collaboration network.}
    \label{table:static}
\end{table}

In another experiment with this dataset, we evaluate the consistency of the sampling method with the Central Limit Theorem. We fixed the number of nodes at 2,500 and conducted experiments with various sample sizes. For this experiment, we only used the \textbf{Uniform Node Sampling (UNS)} method.

\subsection{Temporal Network}
\label{sec:temporal}
\begin{figure}[tbhp]
\centering
\includegraphics[width=1\linewidth]{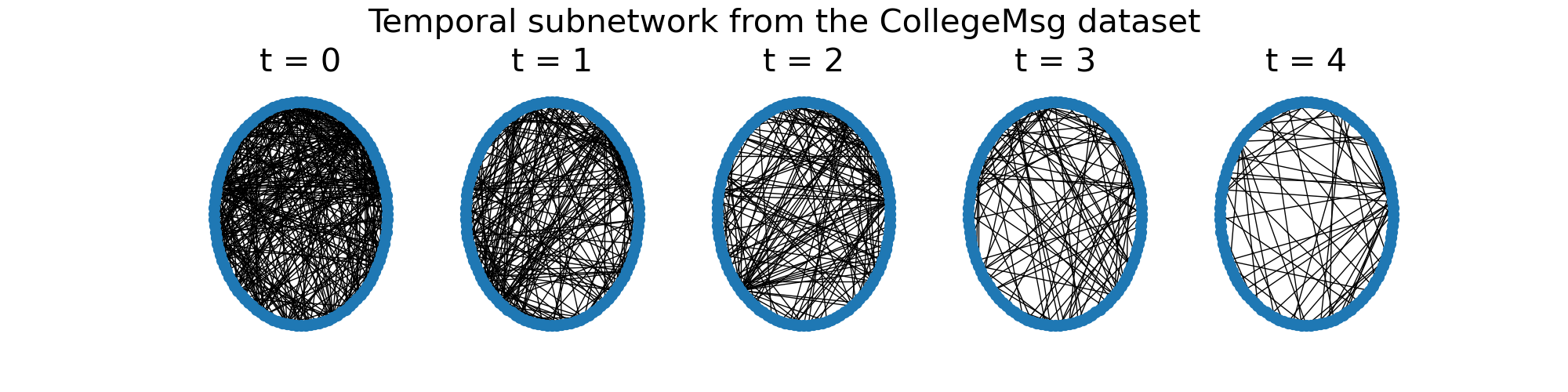}
\caption{In the temporal network for this study, the number of edges decreases over time, suggesting that the relationships are short-term, as individuals are no longer sending messages to each other. The subnetwork consists of 116 nodes, which remain constant (this is a multiplex network). The original network contains 1,899 nodes and 59835 temporal edges, while the static projected graph has 20296 edges. For simplicity, we transform this directed network to an undirected one.
}
\label{fig:dataset_temporal}
\end{figure}

The second dataset we would examine is the CollegeMsg temporal network \cite{10.5555/1543767.1543769}. This dataset comprises private messages exchanged on an online social network at the University of California, Irvine. Users had the ability to search for others within the network and initiate conversations based on profile information. An edge \((u, v, t)\) indicates that user \(u\) sent a private message to user \(v\) at time \(t\). For our analysis, we only focus on a small set of users that remain constant over time and group the timestamps into bins of 40 days. This means that an edge \((u, v, t = n)\) signifies that user u sent a private message to user v during the \( n^{th} \) 40-day period. Thus, this is a multiplex and temporal network. More details about this network can be found in Fig. \ref{fig:dataset_temporal} and Table \ref{table:temporal}.

In this experiment, we fixed the number of nodes in each sampling subnetwork at 30 and varied the sample size. For the sampling, we sampled the nodes in the network at an initial time \( G(t=0) \) using specific methods, and then we extracted the induced subnetworks at every timestep with the same sampled set of nodes. After generating the samples, we will calculate the average degree, the global clustering coefficient, and the edge percentage, and then compare these metrics with those of the original network. The sampling methods used in this experiment include \textbf{Uniform Node Sampling (UNS)} and \textbf{PageRank Sampling (PRS)} \cite{10.1145/1150402.1150479}, in which nodes are selected based on their PageRank centrality.

\begin{table}[tbhp]
    \centering
    \begin{tabular}{|c c c c c|} 
         \hline
         & No. of nodes & No. of egdes & C & \(\left<k\right>\) \\
         \hline
         t = 0 & 116 & 382 & 0.1187 & 6.5862 \\ 
         t = 1 & 116 & 250 & 0.0751 & 4.3103 \\
         t = 2 & 116 & 188 & 0.1065 & 3.2414 \\
         t = 3 & 116 & 119 & 0.0452 & 2.0517 \\
         t = 4 & 116 & 069 & 0.0119 & 1.1897 \\
         \hline
    \end{tabular}
    \caption{\normalfont A detailed analysis of the after-processing network at every snapshot.}
    \label{table:temporal}
\end{table}

\subsection*{Metric}

The metrics utilized in this study are outlined below:

\begin{itemize}
    \item \textbf{Average degree}:
    \begin{align}
        \left<k\right> = \sum_{i} k_i,
    \end{align}
    where \(k_{i}\) is the number of edges connected to node \(i\).
    \item \textbf{Average shortest path}:
    \begin{align}
        a = \frac{1}{n(n-1)} \sum_{i,j(i \neq j)} d_{ij},
    \end{align}
    where \(d_{ij}\) is the shortest path from \(i\) to \(j\).
    \item \textbf{Edge percentage}:
    \begin{align}
        r_E(t) = \frac{|E(t)|}{|E(t=t_0)|},
    \end{align}
    where \(|E(t)|\) is the number of edges at snapshot \(t\).
    \item \textbf{(Average) clustering coefficient}: 
    \begin{align}
        C = \frac{1}{n} \sum_{i} c_i,
    \end{align}
    where \(c_{i}\) is defined by:
    \begin{align}
        c_{i} = \frac{\text{\# of pairs of neighbors of node i that are connecte}}{\text{\# of pairs of neighbors of node i}}
    \end{align}
    \item \textbf{s-metric} \cite{li2005theoryscalefreegraphsdefinition}: is defined as:
    \begin{align}
        s_{metric} = \sum_{u, v} \text{deg}(u) * \text{deg}(v)
    \end{align}
    \item \textbf{Largest component}: the largest group of nodes (S) where every node can be reached from every other node within this component.
\end{itemize}

\section{Results}

Our task is to reassess the statistical significance of the claim that network sampling can extract the properties of a network without requiring a study of the entire large network in the real world. Most importantly, we aim to analyze the statistics of network metrics for comparison. To achieve this, we collected data on various metrics, including average degree, clustering coefficient, largest component size, and average shortest path length, with the size of the sampled subnetworks ranging from hundreds to thousands of nodes. The original networks exhibit complex properties, such as being directed or containing self-loops; however, we treat them as simple, undirected graphs in order to investigate their most fundamental topological structures.

\begin{figure}[htb!] 
    \centering
    \includegraphics[width=\linewidth]{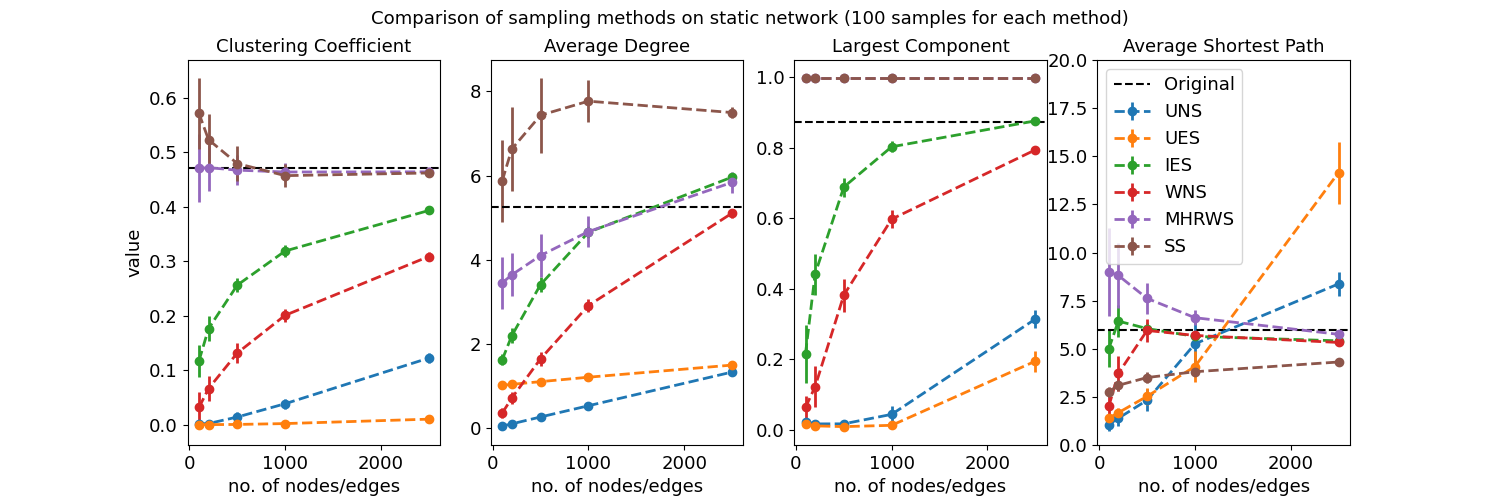}
    \caption{Performance comparison of six sampling methods on a static network across key structural metrics over the CA-HepTh network. The black dashed line represents the values from the original full network. As sample size increases, most methods show convergence toward the true metrics, though the rate and accuracy of convergence vary. Node-based methods effectively approximate node-level properties such as average degree, while edge-based methods better preserve global structures like the largest component size. Exploration-based methods capture connectivity patterns but exhibit bias toward high-centrality nodes. Uniform sampling performs poorly due to its mismatch with the scale-free nature of real-world networks.}.
    \label{fig:comparison_static}
\end{figure}

In the static network, Fig. \ref{fig:comparison_static} presents the statistical results comparing six methods, with further details provided in Section \textbf{Static Network}. The black dashed line indicates the metrics of the original network. As the number of nodes in the sampled network increases, the sampling results begin to converge towards the true values. However, the degree of convergence varies among the methods. Some methods even exhibit divergence, as seen in the plot of the average shortest path, while others show little to no change, depending on the characteristics of the methods (as illustrated in the plot of the largest component). For node-based methods, a proper distribution can effectively capture the node structure, such as the average degree. Similarly, edge-based methods can accurately approximate the size of the largest component, reflecting the edge structure of the network. Exploration-based methods provide good estimates of the connectivity behaviors within networks, but they tend to be biased toward the most critical nodes, much like edge-based methods. In contrast, uniform sampling leads to poor approximations since the distribution of real networks does not conform to a uniform distribution; these networks are typically scale-free.

\begin{figure}[htb!] 
    \centering
    \includegraphics[width=\linewidth]{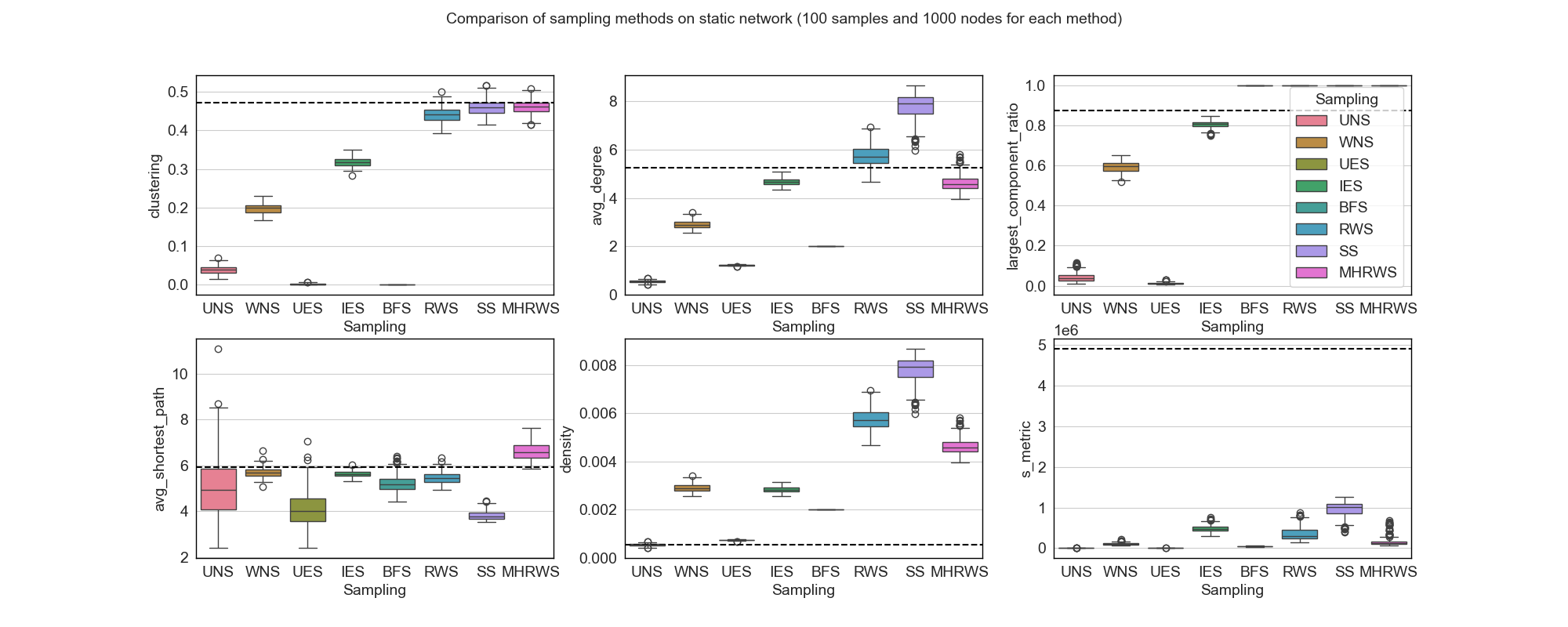}
    \caption{These boxplots compare the performance of eight different network sampling methods on the CA-HepTh network. Each method was evaluated using 100 samples, each consisting of 1,000 nodes. Dashed black lines in the plots represent the corresponding values from the original full network. Results indicate that methods such as RWS, SS, and MHRWS maintain higher fidelity across most metrics, closely approximating the original network structure. In contrast, methods like UNS and UES demonstrate significant deviations from the original structure.}.
    \label{fig:boxplot}
\end{figure}

We evaluated eight sampling methods on a static network by using 100 independent samples, each consisting of 1,000 nodes. The results, summarized in Fig. \ref{fig:boxplot}, illustrate the distribution of key structural properties: average clustering coefficient, average degree, largest component ratio, average shortest path length, density, and s-metric. The dashed black lines in the figure represent the corresponding values from the original network for each metric. The performance of the sampling methods varied significantly in preserving structural features. Exploration-based methods, such as RWS, SS, and MHRWS, consistently approximated the original network across most metrics. Both RWS and SS effectively preserved the clustering coefficient and average degree, achieving the highest ratios of the largest connected component, which indicates strong performance in maintaining global connectivity. MHRWS slightly underestimated the average degree but showed reasonable accuracy in the other metrics. On the other hand, random selection-based methods (UNS, WNS, UES) exhibited considerable discrepancies. UNS and UES significantly underestimated the clustering coefficient and density, resulting in fragmented samples with low largest component ratios. While BFS preserved path-related properties better than UNS and UES, it produced poor clustering values due to its bias towards high-degree nodes. Overall, RWS and SS demonstrated the most robust performance, offering reliable structural fidelity across a range of topological metrics. This makes them well-suited for applications that require representative subgraph extraction from large static networks.

\begin{figure}[htb!] 
    \centering
    \includegraphics[width=\linewidth]{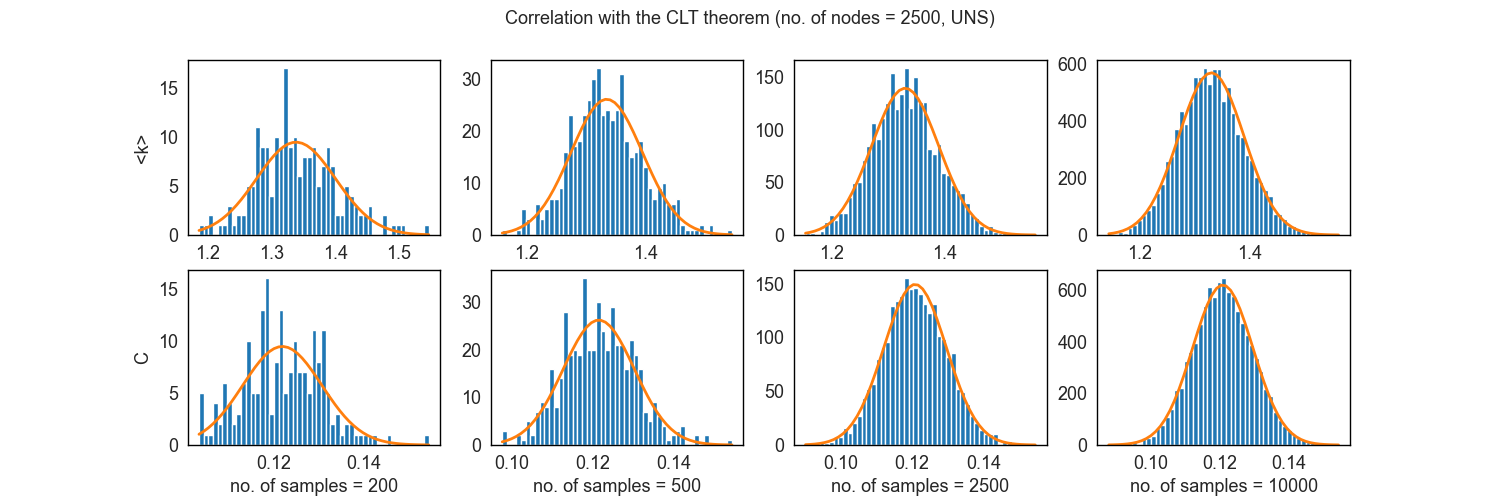}
    \caption{Histograms displaying the average degree (upper plots) and global clustering (lower plots) are compared to the normal distribution, represented by the orange line, for the UNS method. While the degree distribution of the original network follows a power law, the metrics derived from the samples generated by the UNS method remain consistent with the Central Limit Theorem (CLT) as we increase the number of samples.}.
    \label{fig:histogram}
\end{figure}

Fig. \ref{fig:histogram} demonstrates that under the Uniform Node Sampling (UNS) method, both the average degree and global clustering metrics exhibit distributions that closely resemble a normal distribution as the number of samples increases. This behavior is consistent with the Central Limit Theorem (CLT), which indicates that, despite the original network's power-law degree distribution, the sampling distribution of these aggregated metrics stabilizes around a mean. These results suggest that UNS can provide statistically reliable estimates of certain global properties, even if it does not effectively preserve structural characteristics at the microscopic level. This further emphasizes the distinction between the accuracy of sampled metric distributions and the preservation of the network's structural heterogeneity.

\begin{figure}[htb!] 
    \centering
    \includegraphics[width=\linewidth]{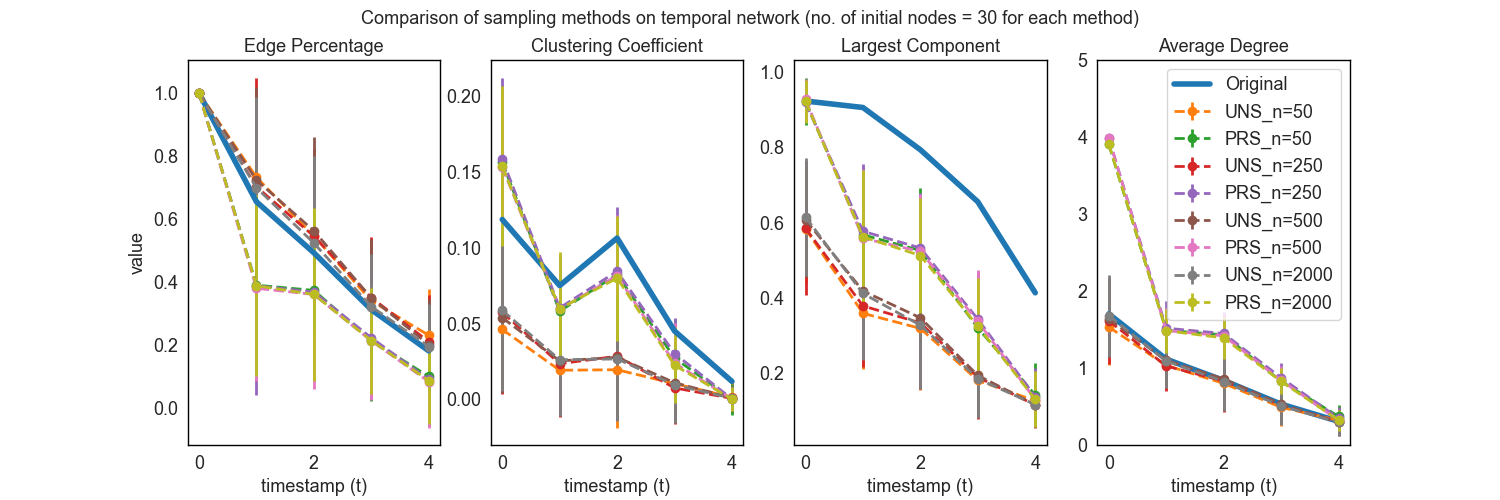}
    \caption{Comparison of sampling methods on a temporal CollegeMsg network with the original value (the blue line), evaluating structural and connectivity metrics. Despite sampling only at the initial timestamp \((t = t_{0})\), sampled subnetworks capture key temporal dynamics. Uniform Node Sampling yields accurate structural (node and edge) estimates but poorly approximates connectivity, whereas PageRank Node Sampling does the reverse. These contrasting results suggest the presence of uniformly random characteristics in the temporal network, warranting further study. Sampling size notably influences the accuracy of metric preservation.}
    \label{fig:comparison_temporal}
\end{figure}

In contrast, for the temporal network, Fig. \ref{fig:comparison_temporal} illustrates the opposite results. Additional details about the experiment can be found in Section \textbf{Temporal Network}. Even when we only sampled at the first timestamp \((t=t_{0})\), the sampled subnetworks still reflected the dynamics of the original network over time. The simple method, Uniform Node Sampling, provides a good estimation of the node and edge structure but offers a poor approximation of connectivity. In contrast, PageRank Node Sampling presents the opposite situation. This raises the question of whether the temporal network exhibits some uniformly random features that require further investigation. Both methods indicate that the size of the sample affects the statistical approximation of the network properties.

\section{Discussion}
This study presents a comparative evaluation of six representative network sampling methods applied to both static and temporal real-world networks. Our findings indicate that no single method consistently outperforms the others across all metrics and network types. In static networks, advanced methods such as Weighted Node Sampling and Random Walk-based strategies offer more accurate approximations of global structural features, particularly for metrics related to connectivity and centrality. In contrast, in temporal networks, simpler methods like Uniform Node Sampling surprisingly provide better structural estimations, while advanced methods tend to introduce bias by oversampling highly active or central nodes.

These results highlight an important insight: \textbf{the effectiveness of a sampling method is highly dependent on the context}, shaped by both the underlying structure of the network and the specific metrics being measured. Additionally, temporal dynamics add further complexity that may not align with the assumptions of traditional sampling strategies designed for static graphs.

The study has several limitations. First, it focuses on only two datasets, which, although representative, may not capture the full diversity of network structures encountered in practice. Second, the evaluation considers a fixed set of metrics; other relevant features such as community structure, resilience, or spreading dynamics were not assessed.

\section{Conclusion}
In this study, we conducted a systematic comparison of six network sampling methods across a static network and a temporal network. Our results demonstrate that no method is universally optimal; the performance of each sampling strategy varies depending on the network type and the metrics under consideration. While advanced methods perform well on static networks, simpler approaches often yield better estimates in temporal settings. These findings highlight the importance of selecting sampling methods based on both network properties and analytical objectives. Future work will expand the scope of datasets and explore metric-specific and application-driven sampling strategies. Moreover, developing adaptive sampling methods that respond to real-time network changes remains a promising direction, particularly for streaming or time-evolving systems.

\section{Acknowledgements}

This research did not receive funding from any public, commercial, or non-profit organization.

\section{Data availability} All datasets are publicly available and downloadable at the Stanford Network Analysis Project at \href{https://snap.stanford.edu/index.html}{https://snap.stanford.edu/index.html}

\bibliographystyle{unsrt}
\bibliography{references}

\begin{thebibliography}{10}

\bibitem{doi:10.1177/0081175012461248}
Ted Mouw and Ashton~M. Verdery.
\newblock Network sampling with memory: A proposal for more efficient sampling from social networks.
\newblock {\em Sociological Methodology}, 42(1):206--256, 2012.
\newblock PMID: 24159246.

\bibitem{AIROLDI2011506}
Edoardo~M. Airoldi, Xue Bai, and Kathleen~M. Carley.
\newblock Network sampling and classification: An investigation of network model representations.
\newblock {\em Decision Support Systems}, 51(3):506--518, 2011.

\bibitem{NBERw25270}
Chih-Sheng Hsieh, Stanley I.~M Ko, Jaromír Kovářík, and Trevon Logan.
\newblock Non-randomly sampled networks: Biases and corrections.
\newblock Working Paper 25270, National Bureau of Economic Research, November 2018.

\bibitem{https://doi.org/10.1002/mpr.2034}
Giovanni Briganti, Marco Scutari, Sacha Epskamp, Denny Borsboom, Ria H.~A. Hoekstra, Hudson~Fernandes Golino, Alexander~P. Christensen, Yannick Morvan, Omid~V. Ebrahimi, Giulio Costantini, Alexandre Heeren, Jill~de Ron, Laura~F. Bringmann, Karoline Huth, Jonas M.~B. Haslbeck, Adela-Maria Isvoranu, Maarten Marsman, Tessa Blanken, Allison Gilbert, Teague~Rhine Henry, Eiko~I. Fried, and Richard~J. McNally.
\newblock Network analysis: An overview for mental health research.
\newblock {\em International Journal of Methods in Psychiatric Research}, 33(4):e2034, 2024.

\bibitem{s21051905}
Jaekoo Lee, MyungKeun Yoon, and Song Noh.
\newblock Advanced network sampling with heterogeneous multiple chains.
\newblock {\em Sensors}, 21(5), 2021.

\bibitem{Ahmed2011NetworkSV}
Nesreen Ahmed, Jennifer Neville, and Ramana~Rao Kompella.
\newblock Network sampling via edge-based node selection with graph induction.
\newblock 2011.

\bibitem{10.1145/3488560.3498383}
Omri Ben-Eliezer, Talya Eden, Joel Oren, and Dimitris Fotakis.
\newblock Sampling multiple nodes in large networks: Beyond random walks.
\newblock In {\em Proceedings of the Fifteenth ACM International Conference on Web Search and Data Mining}, WSDM '22, page 37–47, New York, NY, USA, 2022. Association for Computing Machinery.

\bibitem{MyakushinaExploringST}
Anna Myakushina and Alex Iosevich.
\newblock Exploring sampling techniques in large graphs and networks.
\newblock 2023.

\bibitem{doi:10.1073/pnas.0501179102}
Michael P.~H. Stumpf, Carsten Wiuf, and Robert~M. May.
\newblock Subnets of scale-free networks are not scale-free: Sampling properties of networks.
\newblock {\em Proceedings of the National Academy of Sciences}, 102(12):4221--4224, 2005.

\bibitem{PhysRevE.64.046135}
Lada~A. Adamic, Rajan~M. Lukose, Amit~R. Puniyani, and Bernardo~A. Huberman.
\newblock Search in power-law networks.
\newblock {\em Phys. Rev. E}, 64:046135, Sep 2001.

\bibitem{10.1145/1150402.1150479}
Jure Leskovec and Christos Faloutsos.
\newblock Sampling from large graphs.
\newblock In {\em Proceedings of the 12th ACM SIGKDD International Conference on Knowledge Discovery and Data Mining}, KDD '06, page 631–636, New York, NY, USA, 2006. Association for Computing Machinery.

\bibitem{JMLR:v22:18-240}
Can~M. Le.
\newblock Edge sampling using local network information.
\newblock {\em Journal of Machine Learning Research}, 22(88):1--29, 2021.

\bibitem{5961350}
Tianyi Wang, Yang Chen, Zengbin Zhang, Tianyin Xu, Long Jin, Pan Hui, Beixing Deng, and Xing Li.
\newblock Understanding graph sampling algorithms for social network analysis.
\newblock In {\em 2011 31st International Conference on Distributed Computing Systems Workshops}, pages 123--128, 2011.

\bibitem{10192005}
Le~Fang and Chuan Wu.
\newblock Hes: Edge sampling for heterogeneous graphs.
\newblock In {\em 2023 International Joint Conference on Neural Networks (IJCNN)}, pages 1--8, 2023.

\bibitem{Jiao2024}
Bo~Jiao.
\newblock Sampling unknown large networks restricted by low sampling rates.
\newblock {\em Scientific Reports}, 14(1):13340, Jun 2024.

\bibitem{10.1007/11422778_27}
V.~Krishnamurthy, M.~Faloutsos, M.~Chrobak, L.~Lao, J.~H. Cui, and A.~G. Percus.
\newblock Reducing large internet topologies for faster simulations.
\newblock In {\em NETWORKING 2005. Networking Technologies, Services, and Protocols; Performance of Computer and Communication Networks; Mobile and Wireless Communications Systems}, pages 328--341, Berlin, Heidelberg, 2005. Springer Berlin Heidelberg.

\bibitem{10.1145/2601438}
Nesreen~K. Ahmed, Jennifer Neville, and Ramana Kompella.
\newblock Network sampling: From static to streaming graphs.
\newblock {\em ACM Trans. Knowl. Discov. Data}, 8(2), June 2013.

\bibitem{5462078}
Minas Gjoka, Maciej Kurant, Carter~T. Butts, and Athina Markopoulou.
\newblock Walking in facebook: A case study of unbiased sampling of osns.
\newblock In {\em 2010 Proceedings IEEE INFOCOM}, pages 1--9, 2010.

\bibitem{4781123}
Christian Hübler, Hans-Peter Kriegel, Karsten Borgwardt, and Zoubin Ghahramani.
\newblock Metropolis algorithms for representative subgraph sampling.
\newblock In {\em 2008 Eighth IEEE International Conference on Data Mining}, pages 283--292, 2008.

\bibitem{Lawler1999}
Gregory~F. Lawler.
\newblock {\em Loop-Erased Random Walk}, pages 197--217.
\newblock Birkh{\"a}user Boston, Boston, MA, 1999.

\bibitem{10.1214/aoms/1177705148}
Leo~A. Goodman.
\newblock {Snowball Sampling}.
\newblock {\em The Annals of Mathematical Statistics}, 32(1):148 -- 170, 1961.

\bibitem{10.1145/1879141.1879192}
Bruno Ribeiro and Don Towsley.
\newblock Estimating and sampling graphs with multidimensional random walks.
\newblock In {\em Proceedings of the 10th ACM SIGCOMM Conference on Internet Measurement}, IMC '10, page 390–403, New York, NY, USA, 2010. Association for Computing Machinery.

\bibitem{10.1145/2505515.2505618}
Meng Fang, Jie Yin, and Xingquan Zhu.
\newblock Active exploration: simultaneous sampling and labeling for large graphs.
\newblock In {\em Proceedings of the 22nd ACM International Conference on Information \& Knowledge Management}, CIKM '13, page 829–834, New York, NY, USA, 2013. Association for Computing Machinery.

\bibitem{BLAGUS2017136}
Neli Blagus, Lovro Šubelj, and Marko Bajec.
\newblock Empirical comparison of network sampling: How to choose the most appropriate method?
\newblock {\em Physica A: Statistical Mechanics and its Applications}, 477:136--148, 2017.

\bibitem{Touwen2024}
Lourens Touwen, Doina Bucur, Remco van~der Hofstad, Alessandro Garavaglia, and Nelly Litvak.
\newblock Learning the mechanisms of network growth.
\newblock {\em Scientific Reports}, 14(1):11866, May 2024.

\bibitem{doi:10.1073/pnas.2018994118}
Muhua Zheng, Guillermo García-Pérez, Marián Boguñá, and M.~Ángeles Serrano.
\newblock Scaling up real networks by geometric branching growth.
\newblock {\em Proceedings of the National Academy of Sciences}, 118(21):e2018994118, 2021.

\bibitem{10204402}
Ning Ding, Yehui Tang, Kai Han, Chao Xu, and Yunhe Wang.
\newblock Network expansion for practical training acceleration.
\newblock In {\em 2023 IEEE/CVF Conference on Computer Vision and Pattern Recognition (CVPR)}, pages 20269--20279, 2023.

\bibitem{10.3389/fdata.2022.797584}
Alexander~J. Freund and Philippe~J. Giabbanelli.
\newblock An experimental study on the scalability of recent node centrality metrics in sparse complex networks.
\newblock {\em Frontiers in Big Data}, Volume 5 - 2022, 2022.

\bibitem{doi:10.1098/rspb.2011.1959}
R.~Kanai, B.~Bahrami, R.~Roylance, and G.~Rees.
\newblock Online social network size is reflected in human brain structure.
\newblock {\em Proceedings of the Royal Society B: Biological Sciences}, 279(1732):1327--1334, 2012.

\bibitem{PhysRevE.73.016102}
Sang~Hoon Lee, Pan-Jun Kim, and Hawoong Jeong.
\newblock Statistical properties of sampled networks.
\newblock {\em Phys. Rev. E}, 73:016102, Jan 2006.

\bibitem{Ahmed_Neville_Kompella_2021}
Nesreen Ahmed, Jennifer Neville, and Ramana Kompella.
\newblock Network sampling designs for relational classification.
\newblock {\em Proceedings of the International AAAI Conference on Web and Social Media}, 6(1):383--386, Aug. 2021.

\bibitem{https://doi.org/10.1111/oik.08650}
Anne McLeod, Shawn~J. Leroux, Dominique Gravel, Cindy Chu, Alyssa~R. Cirtwill, Marie-Josée Fortin, Núria Galiana, Timothée Poisot, and Spencer~A. Wood.
\newblock Sampling and asymptotic network properties of spatial multi-trophic networks.
\newblock {\em Oikos}, 130(12):2250--2259, 2021.

\bibitem{Smith2013}
Jeffrey~A. Smith and James Moody.
\newblock Structural effects of network sampling coverage i: Nodes missing at random.
\newblock {\em Social Networks}, 35(4):652--668, Oct 2013.

\bibitem{Zhang2025}
Ruochen Zhang.
\newblock Optimization of multiple sampling for solving network boundary specification problem.
\newblock {\em Scientific Reports}, 15(1):4221, Feb 2025.

\bibitem{10.1145/1217299.1217301}
Jure Leskovec, Jon Kleinberg, and Christos Faloutsos.
\newblock Graph evolution: Densification and shrinking diameters.
\newblock {\em ACM Trans. Knowl. Discov. Data}, 1(1):2–es, March 2007.

\bibitem{10.5555/1543767.1543769}
Pietro Panzarasa, Tore Opsahl, and Kathleen~M. Carley.
\newblock Patterns and dynamics of users' behavior and interaction: Network analysis of an online community.
\newblock {\em J. Am. Soc. Inf. Sci. Technol.}, 60(5):911–932, May 2009.

\bibitem{li2005theoryscalefreegraphsdefinition}
Lun Li, David Alderson, Reiko Tanaka, John~C. Doyle, and Walter Willinger.
\newblock Towards a theory of scale-free graphs: Definition, properties, and implications (extended version), 2005.

\end{thebibliography}






\end{document}